\documentclass[twocolumn]{aastex631}
\usepackage{amsmath}
\usepackage{xcolor} 

\begin{document}

\title{Recurrent eruptions from the emergence of a toroidal flux tube}

\correspondingauthor{J. Zhuleku}
\email{j.zhuleku@uoi.gr}

\author{J. Zhuleku}
\affiliation{Physics Department, University of Ioannina, Ioannina 45110,
Greece}

\author{V. Archontis}
\affiliation{Physics Department, University of Ioannina, Ioannina 45110,
Greece}

\author{K. Moraitis}
\affiliation{Physics Department, University of Ioannina, Ioannina 45110,
Greece}






\begin{abstract}
%
%
%
%
%
%
%

Solar eruptive behavior is often modeled with magnetohydrodynamic simulations of magnetic flux emergence. The usual geometry considered is that of a horizontal cylindrical magnetic flux tube. An alternative is the toroidal tube geometry which has some advantages over the cylindrical one, namely, that the emerging bipolar pair of sunspots do not drift apart indefinitely. In addition to the toroidal tube, we include an oblique, ambient field in the simulation, which leads to the production of increased activity from the interaction of ambient and emerging fields. Letting the simulation run for a long time reveals that six eruptive jets take place after an initial reconnection jet. In an attempt to better understand the eruptive activity we examine the evolution of free energy and relative helicity of the coronal volume, as well as, the magnetic tension forces acting on the magnetic system. We find that all quantities decrease in magnitude during the eruptive jets and rebuild afterwards. This recurrent activity continues even after flux emergence ceases and stops once the examined quantities saturate to nearly constant values.

\end{abstract}



\section{Introduction} \label{sec:intro}

\begin{figure}
  \hspace{-0.1 cm}
  \begin{minipage}{\columnwidth} 
  \hspace{-1.3 cm}
    \includegraphics[height=6.0 cm, width=10 cm]{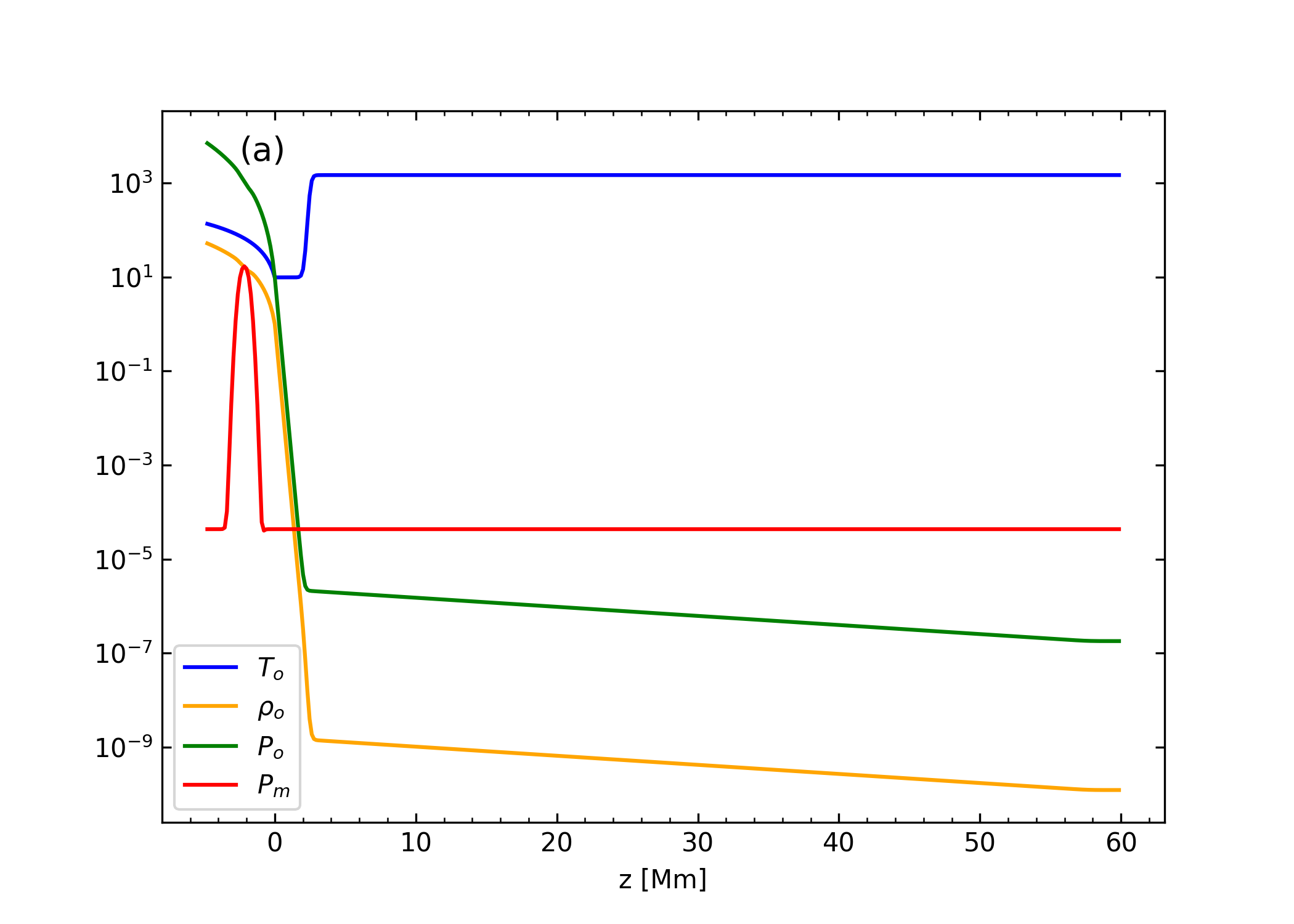}
    \hspace{-0.1 cm}
    \includegraphics[height=5.5 cm, width=7.8 cm]{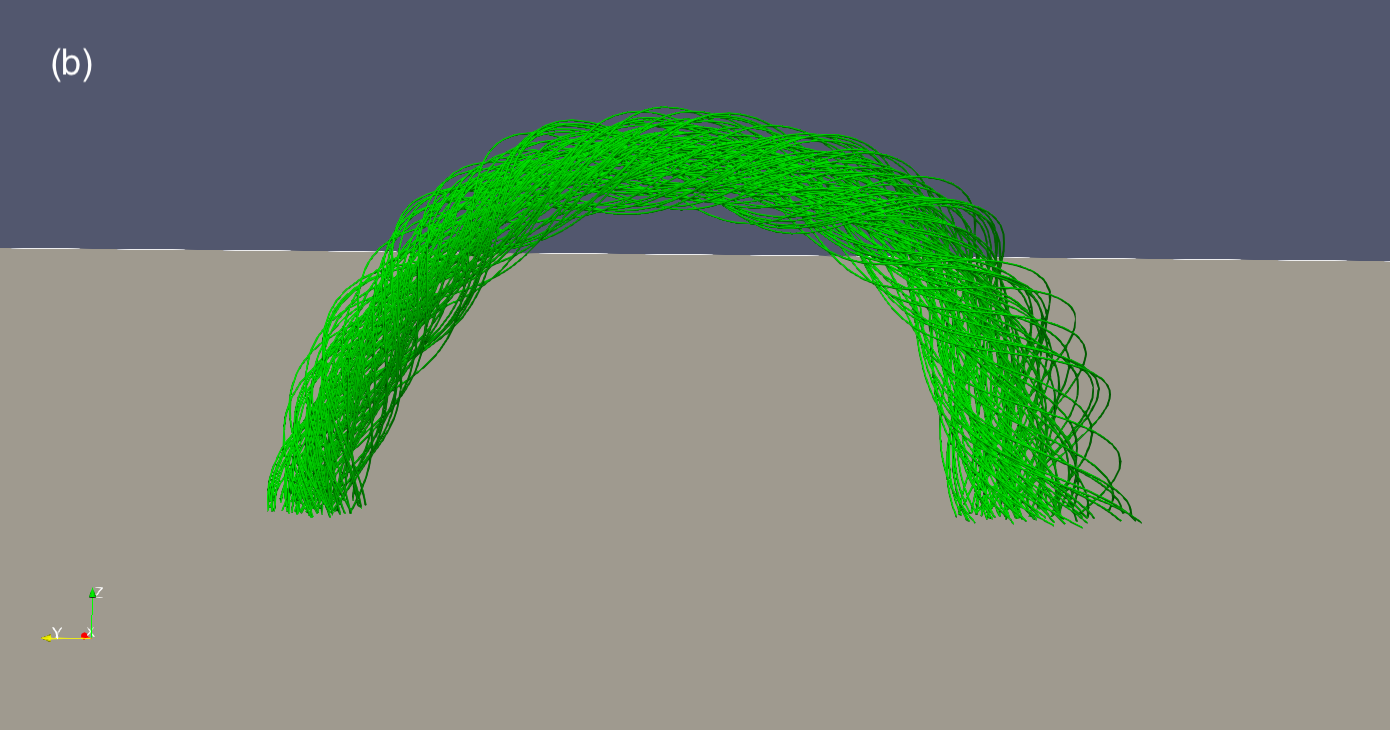}
     \end{minipage}

  \caption{Initial conditions of the numerical experiment: a) The plane parallel stratification of the atmosphere in the box. The plot shows the distribution of temperature (blue), density (orange), pressure (green), and magnetic pressure (red) spanning from the convection zone to the corona. b) A 3D representation of the toroidal tube placed at the interior of the numerical box.}
  \label{fig1}
\end{figure}

Eruptive phenomena in the Sun such as flares, jets, and coronal mass ejections (CME) may result from the interaction between emerging magnetic fields and pre-existing ambient fields and they can be observed in various regions in the Sun, such as, coronal holes, active regions, etc.  
Specifically, solar jets are collimated plasma flows typically observed at X-ray and extreme ultra violet (EUV).
First observations by Yohkoh \citep{Shibata1992} and later by Hinode \citep{Savcheva2007} and STEREO \citep{Patsourakos2008} revealed jets as inverted-Y structures exhibiting lifetimes ranging from 5 to 30~min and velocities typically in the range of 70 to 500~km~s$^{-1}$ \citep{Savcheva2007,Cirtain2007}.
Solar jets can reach temperatures between 3 and 8 MK and densities between $7\times 10^{8}\,\rm{cm}^{-3}$ and $4\times 10^{9}\,\rm{cm}^{-3}$ \citep{Shimojo2000}.

A possible mechanism for the onset and emission of X-ray inverted Y solar jets (hereafter “standard” jets) is magnetic reconnection between oppositely directed emerging and ambient magnetic fields, when they come into contact \citep{Hayverts1997,Forbes1984}.
This interaction leads to the release of stored magnetic energy, which subsequently converts into Joule heating and kinetic energy, accelerating plasma to high velocities.
The concept was initially explored through 2D numerical models by \cite{Yokohama1995,Yokohama1996}.
Their numerical experiments demonstrated the emergence of fast-moving outflows at the reconnection site, resulting in the formation of jets which was documented for the first time in numerical simulations.
Jet formation and the nature of eruption-driven jets has been studied in 3D numerical models by \cite{Insertis2008,Archontis2013,Insertis2013}.
These studies report the formation of a jet consisting of hot and cool plasma followed by a series of energetic eruptions.
These eruptions are commonly referred to as "blowout" jets \citep{Moore2010}.
Triggering such intense eruptions commonly observed in the Sun \citep{Golub2007,Kosugi} has been under debate for decades.
One of the first attempts to understand the physical mechanism behind eruptions is the breakout model\citep{Antiochos1999}.
In this model a flux system is stressed under a boundary imposed shearing of the footpoints leading to a formation of a current sheet.
The occurring reconnection at the current sheet changes the geometrical topology of the field lines resulting in an outward ejection of the flux rope.
Similar results were also reported in the work of \cite{Devore2008}, \cite{Lynch2008}, and \cite{Wyper2017}.
These models emphasize that the formation of a current sheet is a fundamental step for triggering eruptions.
However, most of these studies artificially impose shearing at the polarity inversion line, overlooking the spontaneous, intrinsic processes that might drive solar activity.

The recurrent nature of both large and small-scale solar eruptions has been well documented in the recent years \cite[see review][]{Lugaz2017}.
\cite{Werner2017} report on the observation of the ejection and interaction between multiple CMEs from 6 to 9th of September 2017.
More recently, the newly launched Solar Orbiter observed three CME eruptions from the same region within a short time scale \citep{Mierla2023}.  
Additionally, recurrent "blowout" jets and impulsive flares were observed, originating from the same active region between October 21 and 24, 2003 \citep{Chandra2017} but also recently in 2022 at the NOAA13078 active region as reported by \cite{Zhou2025}.
These studies collectively underscore a common characteristic of energetic phenomena: the capability of a single region to generate multiple eruptive events in time scales well after their emergence.

An important process towards triggering recurrent eruptive phenomena in the Sun is the emergence of the magnetic fields, from the base of the convection zone to the photosphere due to buoyancy \citep{Parker1955}.
In terms of numerical simulations, there is a considerable number of studies, which reported on the emergence of magnetic fields from the solar interior to the outer solar atmosphere \citep{Fan2001,Manchester2004,Archontis2004,Torok2008,Toriumi2011,Toriumi2013,Archontis2013,Syntelis2015,Syntelis2017,Syntelis2019,Choul2023}. 
In the majority of these simulations, a twisted cylindrical horizontal flux tube is embedded in the solar interior, and the introduction of a density deficit results in the formation of an $\Omega$-loop that ascends, giving rise to a small or ephemeral active region. 
When the field emerges at the photosphere, two opposite-polarity sunspots and a strong polarity inversion line (PIL) between them are formed. 
It has been found that the field lines along the PIL are sheared due to the magnetic field \citep{Fan2001}.
As time goes by, the two sunspots move away from each other, along the PIL, indefinitely. 
However, observations of emerging active regions show that sunspots move apart for limited periods of time, and usually they follow a gradual separation until they decay \citep{Liu2006}.
An alternative approach to the cylindrical tube geometry was proposed by \cite{Hood2009} and \cite{Mac2009b}.
In these works the initial condition for the magnetic field has the geometry of a toroidal flux tube with its feet anchored at the base of the computational domain.
This geometry and configuration of the tube result in the emergence of sunspots that gradually drift apart until reaching a maximum separated distance (hereafter SSD), which is determined by the major radius of the toroidal flux tube.
In this paper we start with a similar to \cite{Hood2009} toroidal flux tube as initial condition to a numerical simulation, and investigate the eruptive activity during the separation of the two sunspots at the photosphere, and after the maximum SSD has been reached. In addition, we examine whether there is eruptive activity after flux emergence at the photosphere stops.
To achieve this, we perform a 3D magnetohydrodynamics (MHD) simulation of toroidal flux tube emergence. 
We run the simulation for an extended duration enabling a study of the eruptive activity occurring after the onset of the magnetic flux emergence in the photosphere. 
In section~\ref{sec:model} we present the details of how the flux emergence model is set up, in section~\ref{sec:results} we give the main results of our numerical experiment, and in section~\ref{sec:conc} we discuss the results.


\section{Model setup} \label{sec:model}

We perform our 3D MHD experiment using the Lare3D code \citep{Arber2001}.
Lare3D solves the 3D time-dependent, compressible and resistive MHD equations numerically, in Cartesian geometry.

The MHD equations solved are given in non-dimensional form by:
\begin{equation}
\frac{\partial \rho}{\partial t} +\nabla \cdot (\rho \mathbf{u})=0,
     \label{eq:cont}
\end{equation}

\begin{equation}
\rho\left( \frac{\partial \mathbf{u}}{\partial t}+(\mathbf{u}\cdot \nabla)\mathbf{u}\right)=-\nabla P +\mathbf{j}\times \mathbf{B}+\rho\mathbf{g},
     \label{eq:momentum}
\end{equation}

\begin{equation}
    \frac{\partial \mathbf{B}}{\partial t}=\nabla \times \left(\mathbf{u} \times \mathbf{B}\right)+\eta\nabla^2 \mathbf{B},
    \label{eq:ind}
 \end{equation}

 \begin{equation}
\rho\left( \frac{\partial \epsilon}{\partial t}+(\mathbf{u}\cdot \nabla)\epsilon\right)=-P\nabla \cdot \mathbf{u}+\eta \mathbf{j}^2.
     \label{eq:energy}
\end{equation}
Here $\rho$, $P$, $\mathbf{u}$ and $\mathbf{B}$ are the mass density, pressure, velocity, and magnetic field respectively.
The equations are normalized using the photospheric value of density, $\rho_{\rm{ph}}=1.67\times 10^{-4} \ \rm{kg}$~m$^{-3}$, the magnetic field strength $B=300$~G, and the length $H=180$~km. From these follow the units of velocity, $u=2.1 \ \rm{km}$~s$^{-1}$, temperature, $T=649$~K, and time, $t=86.9$~s.
The specific energy density is calculated as $\epsilon=P/((\gamma-1)\rho)$, with $\gamma=5/3$.
The current density is calculated by Amp\`ere's law, $\mathbf{j}=\nabla \times \mathbf{B}$, and gravity is equal to the solar value, $g=274\,\rm{m}\,\rm{s}^{-2}$.
Resistivity $\eta$ is assumed uniform, with a constant value 0.01 in dimensionless units.

The equations are solved inside a $420^3$ Cartesian box similar to those used by \cite{Syntelis2017} and \cite{Choul2023}.
The physical size of the computational box is [$-30$, 30] $\times$ [$-30$, 30] $\times$ [$-4.9$, 60] $\rm{Mm}$ in the $x$, $y$ and $z$ directions.  
The background stratification, which is shown in Figure~\ref{fig1}a, has the structure that is typically used \citep{Fan2001,Archontis2004,Hood2009,Mac2009,Mac2009b}. 
The non-magnetic stratification includes a solar interior ($-4.9 \ \rm{Mm} < z < 0$), a photosphere ($ 0< z < 1.8 \ \rm{Mm}$), a chromosphere/transition region ($1.8 \ \rm{Mm} < z < 3.2 \ \rm{Mm}$), and a corona ($ 3.2 \ \rm{Mm} < z < 60 \ \rm{Mm}$).
The boundary conditions are open on the top and $x$ direction, closed at the base of the computational box, and periodic on $y$ direction.

Following the works of \cite{Hood2009} and \cite{Mac2009}, a toroidal tube is placed initially inside the convection zone, as shown in Figure~\ref{fig1}b. The magnetic field components of the tube are expressed as
\begin{equation}
    B_x = B_{\rm{\theta}}(r)\frac{s-s_0}{r},
\end{equation}
\begin{equation}
    B_y = -B_{\rm{\phi}}(r)\frac{z-z_0}{s}- B_{\rm{\theta}}(r)\frac{x}{r}\frac{y}{s}, 
\end{equation}
\begin{equation}
    B_z = B_{\rm{\phi}}(r)\frac{y}{s}- B_{\rm{\theta}}(r)\frac{x}{r}\frac{z-z_0}{s},
\end{equation}
where
\begin{equation}
    r^2=x^2+(s-s_0)^2, \quad s-s_0=r \cos \rm{\theta}, \quad x=r \sin \rm{\theta},
\end{equation}
and
\begin{equation}
    B_{\rm{\phi}}=B_0e^{-r^2/r^2_0}, \quad B_{\rm{\theta}}=\alpha rB_{\rm{\phi}}=\alpha rB_0e^{-r^2/r^2_0}.
\end{equation}
Here, $s_0$ corresponds to the major radius of the tube, $r_0$ to the minor radius, $z_0$ to the base of the convection zone, $B_0$ to the initial magnetic field strength of the tube, and $\alpha$ to its initial twist.
The parametric study of \cite{Mac2009} showed that the variation of the various parameters of the toroidal flux tube such as the magnetic field $B_0$ or the twist $\alpha$ can have a significant impact on the overall dynamic properties of the emerging field.
In this numerical experiment we keep these values fixed, with the major radius of the tube set to $s_0=15$ (2.7~Mm), the minor radius to $r_0=2.5$ (0.45~Mm), magnetic field strength $B_0=21$ (6300~G), twist $\alpha=0.4$, and $z_0=27$ (4.9~Mm).
In addition to the toroidal tube, a uniform oblique ambient field is included with an angle between the field lines and the $z$-axis equal to $\theta=11^{o}$, and with the $x$-axis on the horizontal plane equal to $\phi=183^{o}$. 
The strength of the uniform field is set to $B_{\rm{amb}}\simeq 0.033$ (10~G).

\begin{figure*}
\hspace{-2.5 cm}
\begin{minipage}[t]{\textwidth}
\includegraphics[height=4 cm, width=1.25\linewidth]{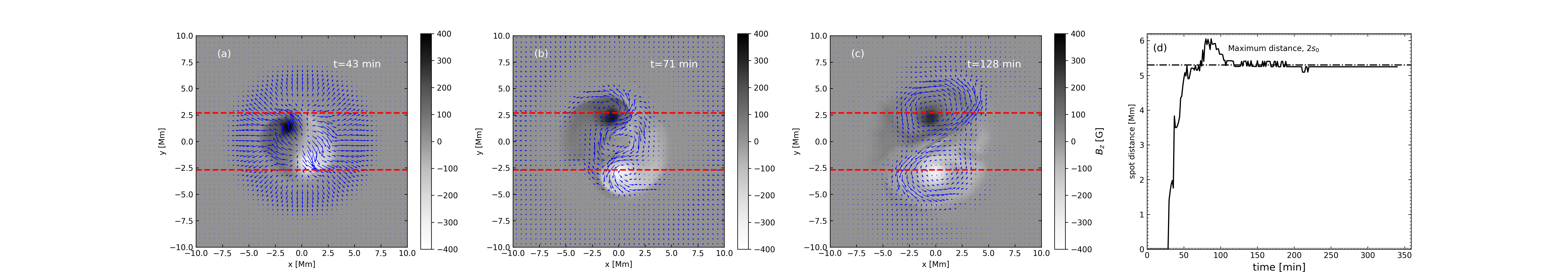}
\end{minipage}
\caption{Evolution of the two sunspots after emergence. a) The $B_z$ components measured at the photosphere for $t=43$~min. b) same for $t=71$~min. c) same for $t=128$~min. Red dashed lines indicate the maximum distance sunspots can reach determined by the major radius of the flux tube. d) The distance of the two polarities as a function of time. Black dashed line indicates the maximum distance $2s_0$. Blue arrows in a)-c) show the horizontal velocity components, $u_{x}$ and $u_{y}$.
}
\label{fig2}
\end{figure*}

\section{Results} \label{sec:results}

The MHD simulation with the setup described in section~\ref{sec:model} was run for a long time, $\sim 250$ snapshots, corresponding to $\sim 350$~min of real time. We describe in the following the phenomena that took place during this time.

\subsection{Sunspots distance over time}
An advantage of using a toroidal tube is that the formed sunspots attain a maximum separation distance. We examine here whether this holds in our case where an ambient field is also present. 
In Figure~\ref{fig2}, we present snapshots of the vertical magnetic field, $B_z$ on the photosphere at three time instances: $t=43$~min, $t=71$~min, and $t=128$~min. 
Following the emergence at $t=43$~min, the two sunspots gradually drift apart until they reach their maximum distance, which is indicated by the two red horizontal lines in Figure~\ref{fig2}. This distance corresponds to the diameter of the initial torus.
After that time, the distance between the two polarities remains relatively constant, as shown in the snapshots for $t=71$~min and $t=128$~min.
The velocity as indicated by the blue arrows is radially outwards from the emergence site, weakening as the sunspots move to their maximum distance.
A counterclockwise rotational motion is also evident around the sunspots, consistent with the expansion and untwisting of magnetic field lines during emergence \citep{archontis2015}.
As depicted in Figure~\ref{fig2}, the values of the photospheric magnetic field in the simulation are around 500~G, which are characteristic of small active regions or ephemeral regions on the Sun.

\begin{figure*}
  \begin{minipage}[t]{\textwidth}
    \hspace{-2 cm}    \includegraphics[height=4cm, width=1.2\linewidth]{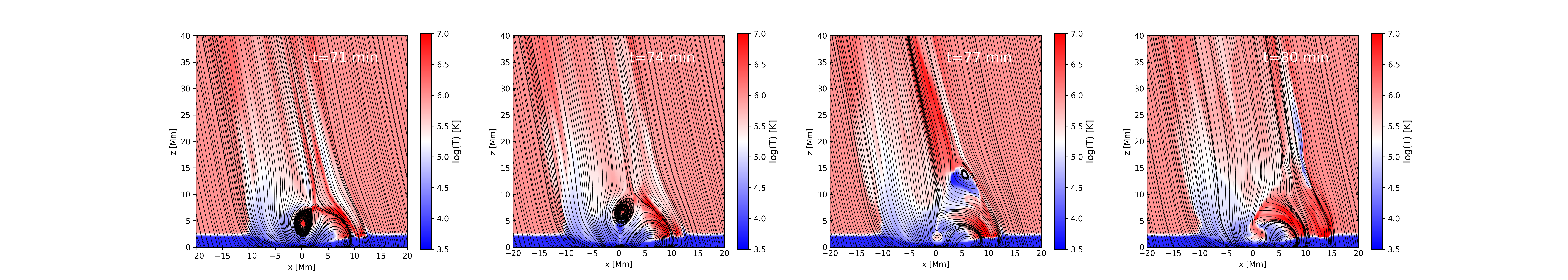}
  \end{minipage}  
  \vspace{0.3cm} 
  \begin{minipage}[t]{\textwidth}
    \hspace{-2 cm}    \includegraphics[height=4cm, width=1.2\linewidth]{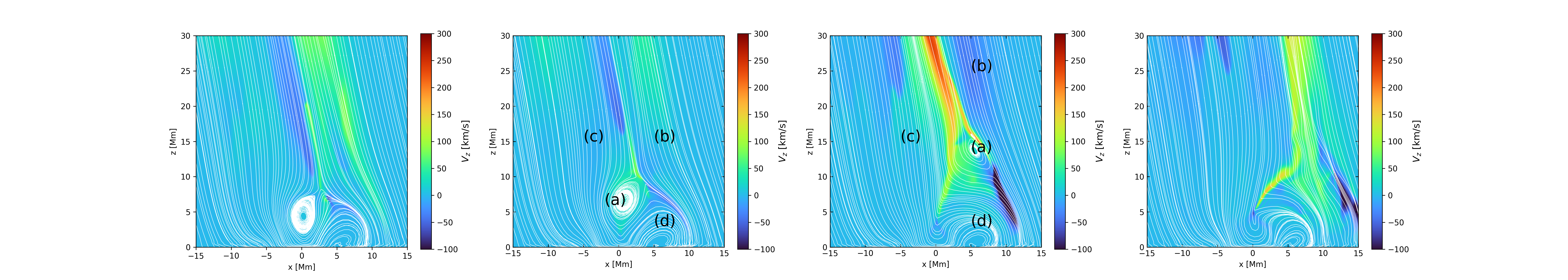}
  \end{minipage}
  \vspace{-0.8 cm}
  \caption{Vertical $x-z$ mid planes ($y=0$) of temperature (top) and velocity (bottom). Top panel shows the magnetic field lines (black) and the formation of the flux rope at times $t=71$~min, $t=74$~min, $t=77$~min, and $t=80$~min. The background represents the temperature. Bottom panel shows the magnetic field lines (white) for the same snapshots but with the vertical velocity in the background. Points (a), (b), (c), (d) are used to trace 3D magnetic field lines. }
  \label{fig3}
\end{figure*}

Initially, at $t=0$, the major radius of the tube is practically the distance between the maximum value of $B_z$ at the one footpoint of the torus and the minimum $B_z$ at the other footpoint.
Therefore, to accurately calculate the separation, we track the centers of the two opposite polarities over time, locating the point with the maximum vertical component of the magnetic field, $B_z$ for the positive polarity ($\rm{x}_{\rm{max}}, \rm{y}_{\rm{max}}$) and the minimum $B_z$ for the negative ($\rm{x}_{\rm{min}}, \rm{y}_{\rm{min}}$). 
The distance between the sunspots is computed then as $d=\left[(\rm{x}_{\rm{max}}-\rm{x}_{\rm{min}})^2 + (\rm{y}_{\rm{max}}-\rm{y}_{\rm{min}})^2 \right]^{1/2}$, and we plot it as a function of time in Figure~\ref{fig2}d. We find that after $t=75$~min the maximum SSD has been reached, although some small oscillations continue up to the end of the simulation.

\begin{figure}
  \begin{minipage}[t]{\textwidth}
    \hspace{-0.1 cm}    \includegraphics[height=4cm, width=0.4\textwidth]{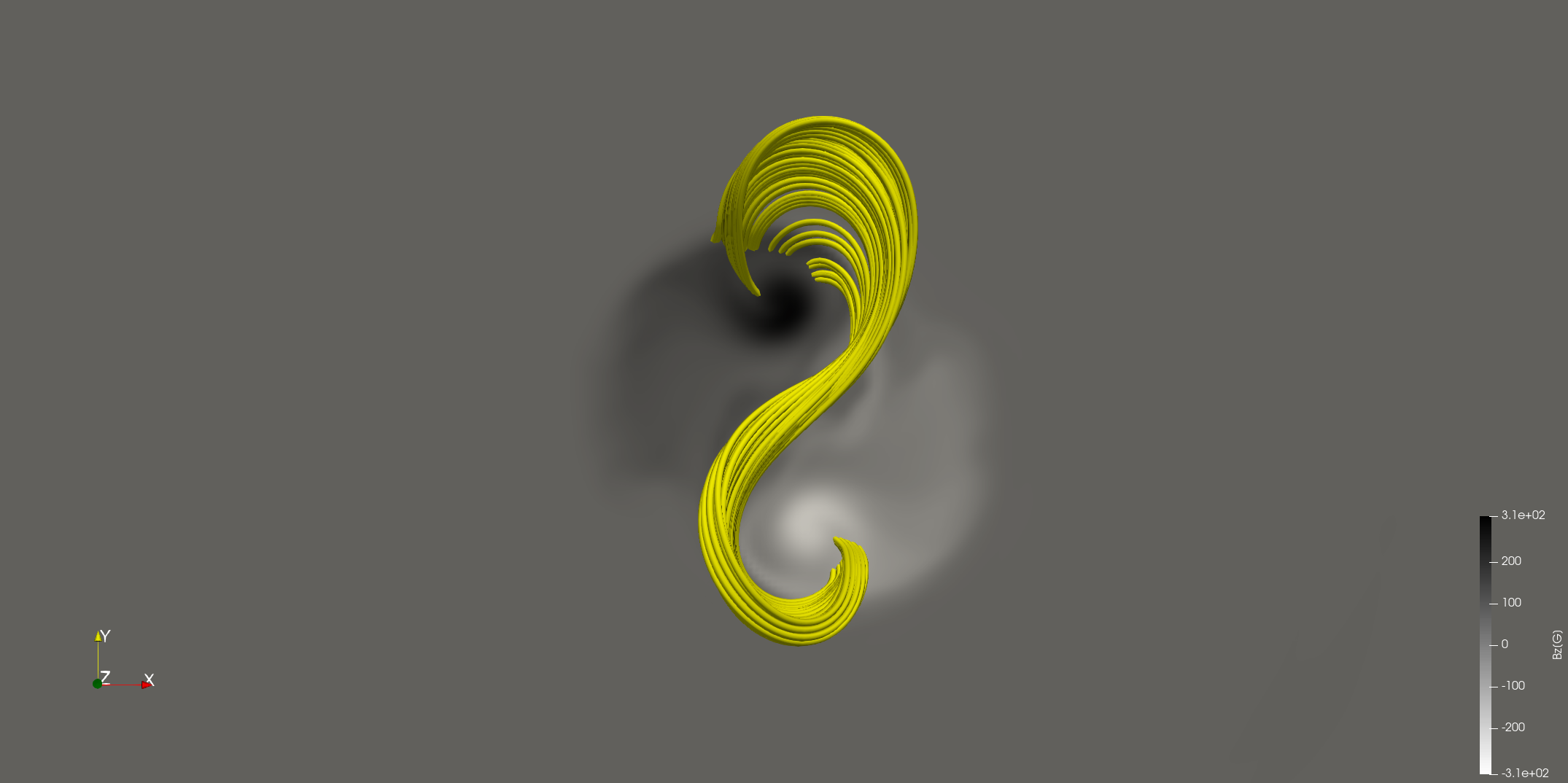}
  \end{minipage}  
  \begin{minipage}[t]{\textwidth}
    \hspace{-0.1 cm}    \includegraphics[height=4cm, width=0.4\textwidth]{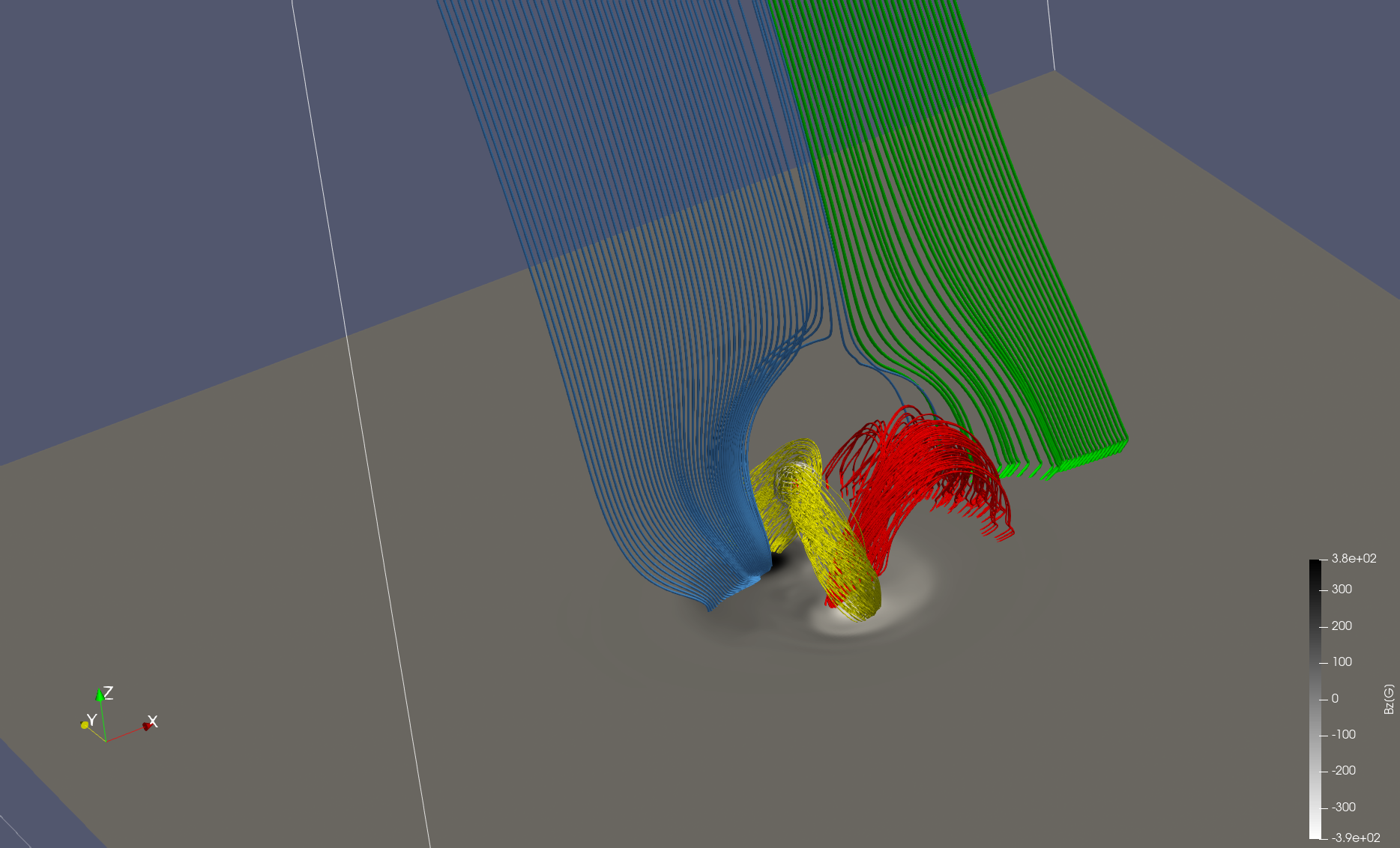}
  \end{minipage}
  \begin{minipage}[t]{\textwidth}
    \hspace{-0.1 cm}    \includegraphics[height=4cm, width=0.4\textwidth]{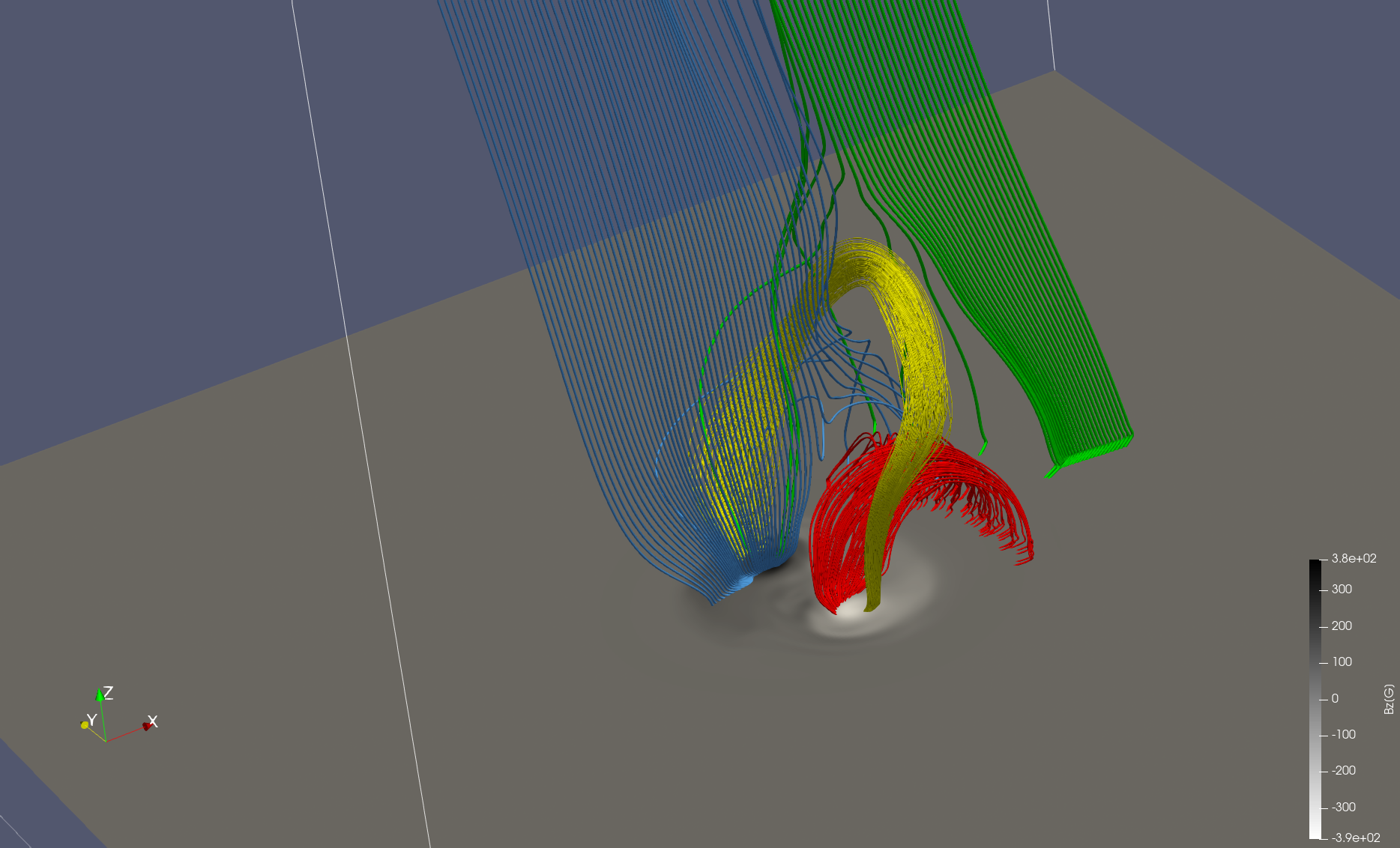}
  \end{minipage}
  \vspace{-0.3 cm}
  \caption{3D representation of the jet structure in our model. Top panel shows the top view of the $B_{z}$ magnetic field at the photosphere along with the flux rope (yellow lines) forming a sigmoid along the PIL at $t=74$~min, right before the first ``blowout" jet. The middle panel shows the 3D structure before the eruption. The green lines represent the non-reconnected ambient magnetic field. The yellow lines represent the magnetic flux rope. The blue lines represent the reconnected field lines. The red lines show the post-flare hot coronal loop formed after the beginning of the ``standard" jet phase. The bottom panel shows the evolution of the magnetic structure after the ``blowout" jet at $t=77$~min. }
  \label{fig4_bl}
\end{figure}

\subsection{Eruption-driven jets}

Including an oblique ambient field in the simulation leads to the formation of a ``standard" reconnection jet, when the ambient field reconnects with the emerging field, as first discussed by \cite{Yokohama1995}.
Similar ``standard" jets have been seen in many numerical models that include horizontal flux tubes \citep[e.g.][]{Archontis2013,Insertis2013}.
After the reconnection jet we find a series of ``blowout" jets which are emitted towards the outer corona. As an example, we show in Figure~\ref{fig3} the onset and evolution of the first ``blowout" jet (and second jet in total), that occurred in the simulation a little after $t=70$~min. 
More precisely, in the first row we show the temperature distribution in the vertical $x$-$z$ mid plane, and in the second row the vertical velocity in the same plane. In both plots a zoom-in of the central region is shown for better visibility.

At $t=71$~min, the ``standard" jet and the post-flare loop underneath are clearly visible, and can be characterized as regions of hot plasma. 
The newly-formed flux rope resides in the central region. 
By $t=74$~min, the flux rope undergoes an eruption, leading to the formation of a ``blowout" jet. 
The ``blowout" jets observed in our study exhibit similarities with a plethora of previous numerical studies \citep[see, e.g.][]{Archontis2013,Insertis2013,Wyper2017}. 
The eruption is accompanied by a significant increase in both the temperature and the width of the jet, as is evident in the snapshots at $t=77$~min and $t=80$~min.
In the bottom panel, we depict the same evolution of the flux rope, but with the vertical velocity shown in the background. 
From the time of the ``blowout" jet, at $t=71$~min, until $t=80$~min, plasma is expelled with velocities reaching approximately 300~km~s$^{-1}$. 
Additionally, at the reconnection site, there is a downward movement of plasma towards lower heights.

In Figure~\ref{fig4_bl} we show a 3D visualization of the magnetic field lines around the ``blowout" jet at two different times, $t=74$~min and $t=77$~min.
In the second and third panel of Figure~\ref{fig4_bl} we show four different sets of field lines traced around the regions marked as (a), (b), (c), (d) in Figure~\ref{fig3}. 
More specifically, the yellow field lines are traced around the center of the flux rope, from region (a).
The top view of the yellow field lines shows that the erupting flux rope which leads to the ``blowout" jet has a sigmoidal shape due to the strong twist.
Furthermore, the green field lines are traced from the non reconnected ambient field lines, from region (b), blue field lines are previously reconnected field lines between emerging and ambient field from region (c), and red field lines represent the post flare loop which is at the base of the Y-inverted shape of the ``standard"-jet picture, from region (d). 
The bottom panel captures the evolution of this structure during the ``blowout" phase at $t=77$~min.
During this phase, the flux rope is being ejected upward along the ambient field lines, disrupting the preexisting magnetic structure.
This eruptive process leads to a significant temperature increase, as demonstrated in Figure~\ref{fig3}.
This is consistent with observations of ``blowout" jets that report temperatures on the order of 10-20~MK emitting mainly in X-ray \citep{Moore2010}.

We should mention that some of the characteristics of the 3D topology during the emission of the ``blowout" jet (such as the sigmoidal erupting flux rope) and the arch-like shape of the post flare loop are similar to the 3D structure of the erupting ``blowout" jets and its surroundings of previous works \citep[e.g.][]{Archontis2013,Insertis2013, Wyper2017} where the mechanism of the eruptions was thoroughly examined.
We anticipate that some of these mechanisms may be at work also in our simulation, despite the different tube geometry that we have as initial condition.
The study of the exact mechanism for the onset of eruptions in our simulation constitutes a separate topic of research, therefore, is out of the scope of this paper and will be addressed in a future work.

\subsection{Evolution of magnetic flux, plasma motions, and eruptive activity}

\begin{figure}
\vspace{-1.1 cm}
\hspace{-0.8 cm}
\includegraphics[height=10 cm, width=8 cm]{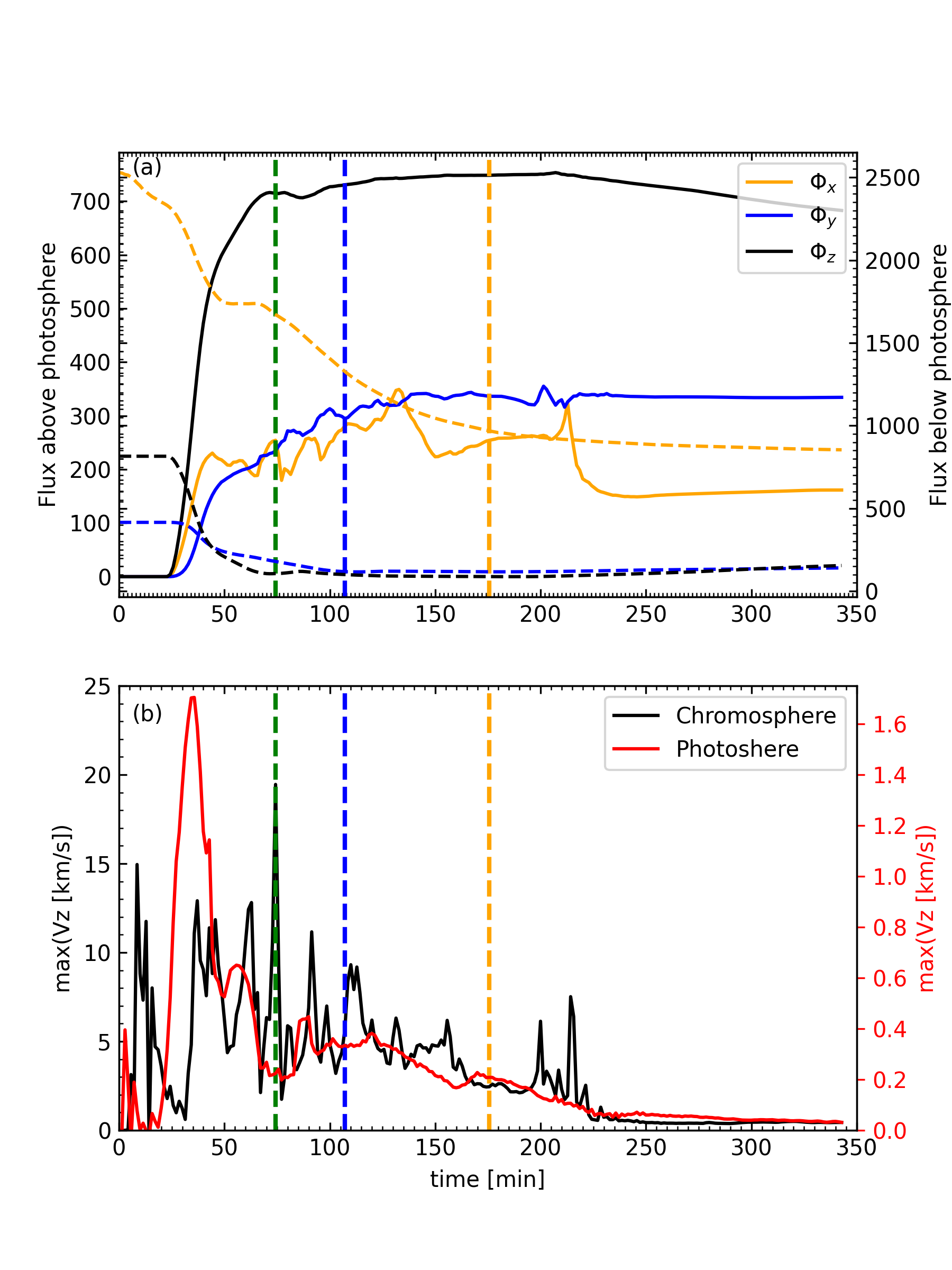}
\vspace{-0.5 cm}
\caption{a) Temporal evolution of the magnetic fluxes $\Phi_{x}$ (orange), $\Phi_{y}$ (blue), $\Phi_{z}$ (black) above (solid lines) and below (dashed lines) the photosphere. b) Temporal evolution of the maximum vertical velocity $V_{z}$ near the base of the photosphere (red line), and at the transition region (black line). The green vertical line marks the time of maximum sunspot separation, the blue vertical line marks the termination of $\Phi_{y}$ and $\Phi_{z}$ emergence, and the orange vertical line marks the termination of $\Phi_{x}$ emergence. 
}
\label{fig_fl}
\end{figure}

To further understand what drives the eruptive phenomena, especially after the maximum SSD is reached, we examine the detailed evolution of magnetic flux components during the simulation.
We start by calculating the magnetic fluxes $\Phi_x$, $\Phi_y$, and $\Phi_z$ defined as
\begin{equation}
    \Phi^\mathrm{above}_x=\int_y \int_{z>0} B_x\, \mathrm{d}y\, \mathrm{d}z, \quad \Phi^\mathrm{below}_x=\int_y \int_{z<0} B_x\, \mathrm{d}y\, \mathrm{d}z,
\end{equation}
\begin{equation}
    \Phi^\mathrm{above}_y=\int_x \int_{z>0} B_y\, \mathrm{d}x\, \mathrm{d}z, \quad \Phi^\mathrm{below}_y=\int_x \int_{z<0} B_y\, \mathrm{d}x\, \mathrm{d}z,
\end{equation}
\begin{equation}
\begin{aligned}
    \Phi^\mathrm{above}_z &=\int_x \int_y B_z\, \mathrm{d}x\, \mathrm{d}y \big |_{z=0\, \mathrm{Mm}}-\int_x \int_y B_z\, \mathrm{d}x\, \mathrm{d}y \big |_{z=60\, \mathrm{Mm}}, \\
    \Phi^\mathrm{below}_z &=\int_x \int_y B_z\, \mathrm{d}x\, \mathrm{d}y \big |_{z=-4.9\, \mathrm{Mm}}-\int_x \int_y B_z\, \mathrm{d}x\, \mathrm{d}y \big |_{z=0\, \mathrm{Mm}} .
    \end{aligned}
    \label{eq:Phz}
\end{equation}
For the $\Phi_{x}$ and $\Phi_{y}$ we calculate the flux above the photoshere, integrating for $z>0$, and for the flux below the photosphere we integrate for $z<0$.
For $\Phi_{z}$, the net vertical flux above and below the photosphere is obtained using equation~\ref{eq:Phz}.
We illustrate the temporal evolution of these fluxes in Figure~\ref{fig_fl}a.
The green dashed line indicates the time ($t\simeq 75$~min) at which the two sunspots reach their maximum separation. 
The axial flux below the photosphere (blue dashed line) is decreasing until $t=110$~min, and then it saturates to a constant value. This means that emergence of $\Phi_y$ above the photosphere takes place up to that time and then it stops until the end of the simulation. 
Similar evolution is seen for the vertical flux below the photosphere ($\Phi_z$; black dashed line).
The azimuthal flux below the photosphere (orange dashed line) shows first a significant decrease until $t=175$~min, and after that time it reaches an almost constant value. This indicates that there is no further emergence of $\Phi_x$ above the photosphere after $t=175$~min.
The fact that $\Phi_x$, $\Phi_y$, $\Phi_z$ remain constant after $t=175$~min indicates that the whole flux emergence process from the solar interior to the solar surface and above stops then.

In addition, we examine the vertical plasma motion at different layers of the atmosphere, both during, and after the flux emergence phase.
In Figure~\ref{fig_fl}b we plot the maximum vertical velocity $V_{z}$ around the polarity inversion line, slightly above the photosphere at $z\simeq 0.15$~Mm (red line), and at the chromosphere at $z\simeq 1.5$~Mm (black line).
During the early phases of flux emergence (until $t<110$~min), $V_{z}$ at the base of the photosphere reaches peak values of $2$~km$\,\rm{s}^{-1}$, consistent with the active upward transport of magnetic flux. 
However, after $t=110$~min, the red line drops significantly, and by $t=200$~min, vertical velocities at the base of the photosphere become negligible, clearly indicating the end of flux emergence.
In contrast, $V_{z}$ at the chromosphere (black line) reaches values of 5–10~km$\,\rm{s}^{-1}$ which persist well beyond $t=200$~min.

\begin{figure}
\vspace{-1.1 cm}
\hspace{-0.8 cm}
\includegraphics[height=7 cm, width=10 cm]{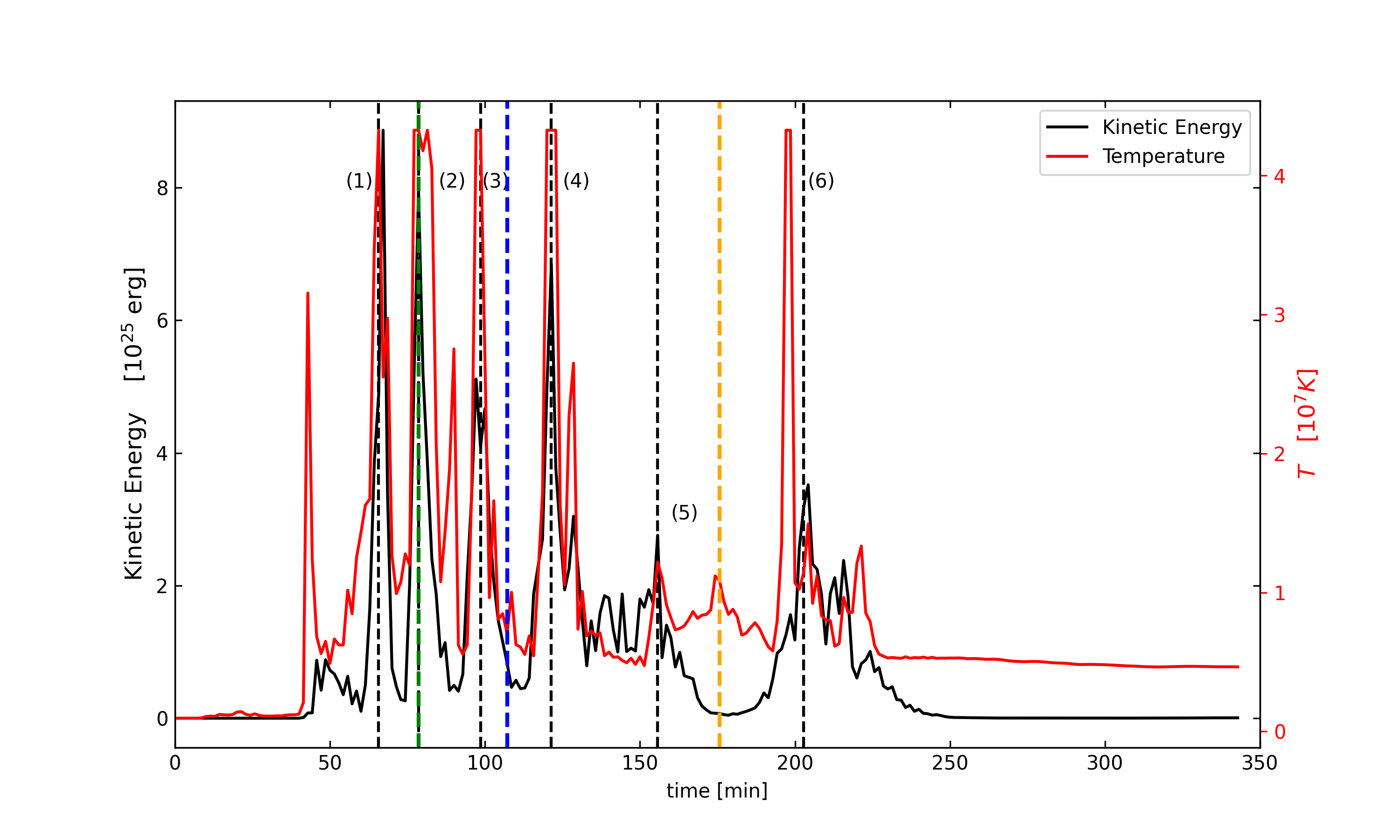}
\vspace{-1. cm}
\caption{Temporal evolution of maximum temperature (red), and kinetic energy (black) at $z=15$~Mm (corona). Peaks in both quantities correspond to eruptive events, labeled (1)–(6) and marked by vertical dashed black lines. The green vertical line indicates the time when the two sunspots reach their maximum separation. The blue vertical line marks the termination of the axial flux emergence, and the orange vertical line marks the termination of all flux emergence.}
\label{fig5_new}
\end{figure}

The series of eruptive events occurs between $t=30$~min and  $t=220$~min.
We plot in Figure~\ref{fig5_new} the temporal evolution of kinetic energy (black line) and plasma temperature (red line) at $z=15$~Mm.
The kinetic energy, which is defined as
\begin{equation}
    K=\iiint \frac{1}{2} \rho u^2\, \mathrm{d}x\, \mathrm{d}y\, \mathrm{d}z,
\end{equation}
is calculated inside a volume centered at $z=15$~Mm, with a thickness of 0.18~Mm along $z$.
The successive peaks correspond to individual eruptions. 
The eruptions are labeled (1)–(6) and are marked by vertical dashed black lines in Figure~\ref{fig5_new}.
The initial temperature peak corresponds to the formation of the ``standard" jet, which is triggered by the reconnection between the emerging and the ambient fields.
This interaction releases less kinetic energy compared to the subsequent ``blowout" jets.
The first five eruptions occur during the flux emergence phase ($t<175$~min) and are associated with significant local maxima in both kinetic energy and temperature. 
These local maxima show strong plasma outflows and intense heating in the corona.
We note here that eruptions (4) and (5) occur after the end of the axial flux $\Phi_{y}$ emergence which is denoted by the blue dashed line.
Later, at $t=200$~min, there is one more eruption (6) which occurs after the whole flux emergence process has stopped.

%

\begin{figure}
\vspace{-1.1 cm}
\hspace{-0.8 cm}
\includegraphics[height=15 cm, width=10 cm]{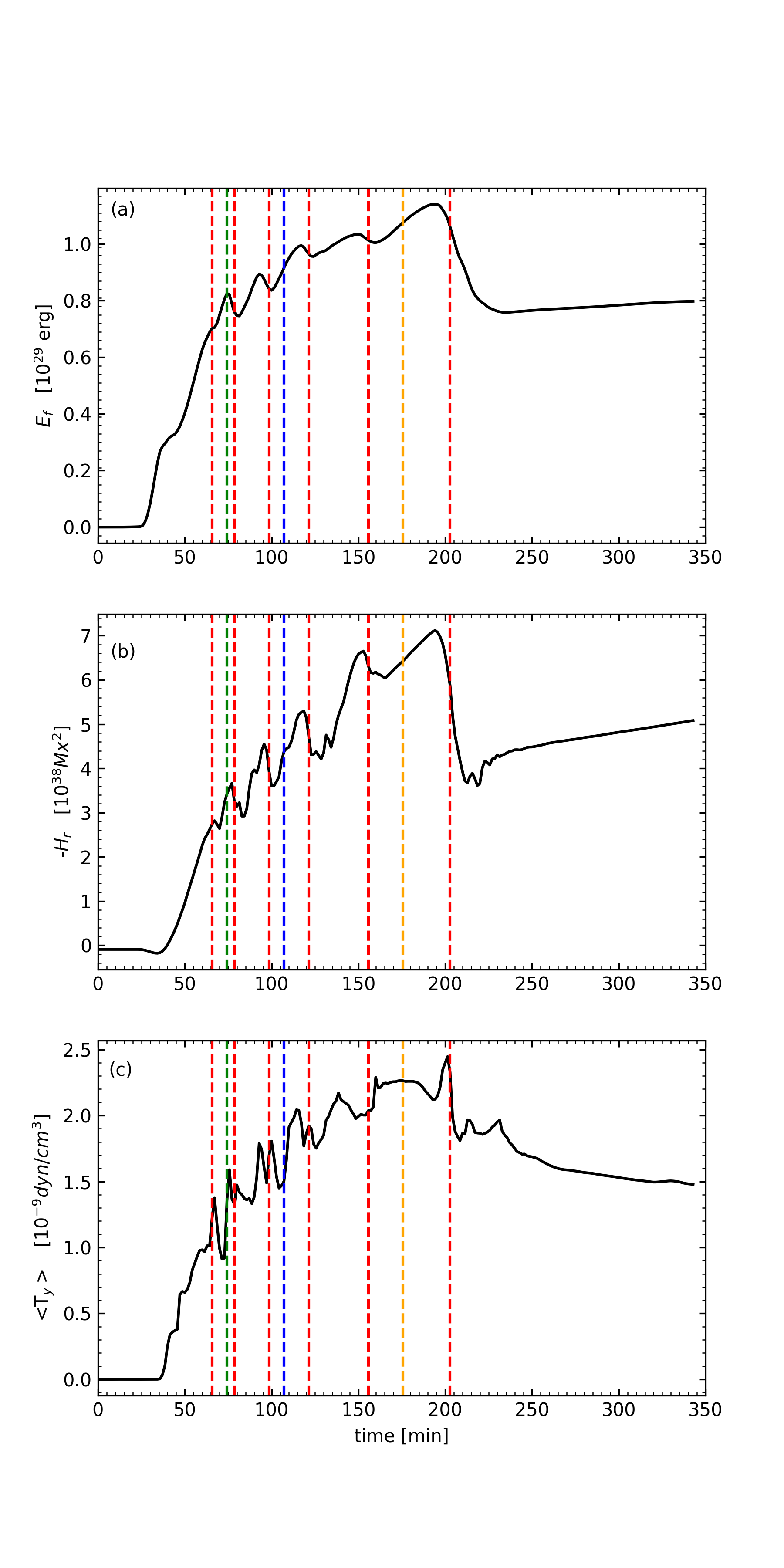}
\vspace{-0.3 cm}
\caption{Evolution of free energy (a), of relative helicity (b), and of average shear force $T_{y}$ for $z>1.2$~Mm (c). Red dashed lines indicate the time of the eruptions. The green dashed line marks the time of maximum sunspot separation, the blue dashed line indicates the termination of axial flux emergence, and the orange dashed line denotes the termination of all flux emergence. 
}
\label{fig_esh}
\end{figure}

\subsection{Evolution of free energy, relative helicity and magnetic tension}
The persistence of eruptive events over an extended period (see Figure \ref{fig5_new}) suggests that energy continues to be released in the corona.
We start by first calculating the free energy in the system.
The free energy ($E_{\rm{f}}$) is calculated as the difference between the total magnetic energy ($E_{\rm{m}}$) and the energy of the potential magnetic field ($E_{\rm{pot}}$),
\begin{equation}
E_{\rm{f}}=\int \frac{B^2}{8\pi}\,\mathrm{d}V - \int \frac{B^2_{\rm{pot}}}{8\pi}\,\mathrm{d}V,
\end{equation}
where $\mathrm{d}V=\mathrm{d}x\,\mathrm{d}y\,\mathrm{d}z$ is the volume element. The potential magnetic field ($\textbf{B}_{\rm{pot}}$) is chosen so as to have the same normal distribution on the boundary of the volume with the MHD field. The computation of such a field requires the numerical solution of a 3D Laplace equation, which is accomplished following the methodology of \citet{moraitis14}. The free energy quantifies the energy available for magnetic reconnection and plasma acceleration, which are indicative of solar eruptions.
In Figure \ref{fig_esh}a we plot the temporal evolution of free energy.
Initially, the free energy increases as the emerging flux and photospheric motions inject energy into the corona.
This increase continues even as the flux emergence slows down.
Peaks in free energy precede eruptions, after which the energy temporarily decreases. 
Following each eruption, $E_{\rm{f}}$ gradually rebuilds before the next eruption, maintaining the recurrent cycle.
However, after the final eruption, the free energy decreases significantly signaling the exhaustion of the system's capacity to sustain eruptive activity.

Another useful diagnostic to analyze eruptions is magnetic helicity.
Magnetic helicity quantifies the degree to which magnetic field lines are twisted and tangle around each other. It is a conserved quantity of ideal MHD and it evolves only through processes like magnetic reconnection. The appropriate form of magnetic helicity in the case of the Sun, where the volume of interest is not magnetically closed, is relative helicity \citep{BergerF84,fa85}. This is defined as
\begin{equation}
    H_{\rm{r}}=\int (\mathbf{A}+\mathbf{A_{\rm{pot}}})\cdot (\mathbf{B}-\mathbf{B_{\rm{pot}}}) \,\mathrm{d}V,
\end{equation}
where $\mathbf{A}$, $\mathbf{A_{\rm{pot}}}$ are the vector potentials of the original and potential magnetic fields, satisfying respectively $\mathbf{B}=\nabla \times \mathbf{A}$ and $\mathbf{B_{\rm{pot}}}=\nabla\times\mathbf{A_{\rm{pot}}}$. Relative helicity is well defined (i.e., gauge-independent) for the potential magnetic field that we use \citep[e.g.,][]{valori16}.
In Figure \ref{fig_esh}b, we plot the temporal evolution of relative helicity ($H_{\rm{r}}$) over the course of the simulation. We notice that relative helicity is negative throughout the simulation in accordance with the sign of the toroidal flux tube's twist.
The behavior of helicity closely follows the behavior of the free magnetic energy shown in Figure \ref{fig_esh}a.
Initially, $H_{\rm{r}}$ steadily increases in absolute value as flux emergence and the motion of the sunspots inject helicity into the corona by stressing and twisting the magnetic field lines.
After $t=75$~min, when the sunspots reach their maximum distance, $H_{\rm{r}}$ continues to grow until the time before the last eruption. 
The peaks of relative helicity follow those of the free energy and correspond to the eruptive events. The magnitude of relative helicity decreases temporarily during each event, as twist is ejected through the upper boundary \citep[e.g.,][]{linan18}.
Similar to the free energy, relative helicity is rebuilt after each eruption, sustaining the sequence of eruptions. 
Relative helicity drops significantly after the final eruption, indicating that the magnetic structures left in the corona are now less helical, in addition to having less free energy, which could explain the termination of the eruptive activity.

To understand the mechanism which drives the eruptions, we also examine the forces acting on the twisted magnetic field lines in the corona.
The force driving these dynamics can be described in a generalized form as the Lorentz force, which can be expressed as follows:
\begin{equation}
    F_{\rm{L}}=-\nabla \bigg (\frac{B^2}{8\pi}\bigg)+\frac{(\mathbf{B}\cdot \nabla)\mathbf{B}}{4\pi}.
\end{equation}
The first term describes the magnetic pressure responsible for the expansion of the magnetic field during its emergence.
The second term corresponds to the tension force.
The $y$-component of this tension force could affect the twisted magnetic field lines within the flux tube.
The tension force is expressed as,
\begin{equation}
T_{y}=\frac{(\mathbf{B}\cdot \nabla)B_y}{4\pi}=\frac{1}{4\pi}\bigg(B_x\frac{\partial B_y}{\partial x}+B_y\frac{\partial B_y}{\partial y}+B_z\frac{\partial B_y}{\partial z}\bigg).
\end{equation}
We calculate the temporal evolution of the average magnitude of this tension force, $<|T_{y}|>$, within the coronal volume above the upper chromosphere (for $z>1.2$~Mm) to focus specifically on the region containing the twisted magnetic structures (Figure \ref{fig_esh}c).
The temporal behavior of $<|T_{y}|>$ is similar to the evolution of the free magnetic energy and relative helicity.
Initially, $<|T_{y}|>$ increases during the flux emergence phase ($t<75$~min), as the motion of the emerging sunspots introduces significant shearing of the twisted magnetic field lines.
After $t=75$~min and until the end of the eruptive activity, $<|T_{y}|>$ continues to increase, 
indicating that the eruptions occurring are primarily magnetically induced, driven by the magnetic tension.
The magnetic tension arises from the curvature and the gradient of the magnetic field lines, essentially described by the term $B\cdot \nabla B_{y}$. 
In regions where $B_{y}$ is strong or exhibits significant spatial variations, the resulting tension force is strong due to the magnetic field lines effectively trying to straighten, or to relieve any bends or twists.
Each peak in $<|T_{y}|>$ corresponds to an eruption, where the release of magnetic tension is linked to reconnection events.
These eruptions partially dissipate the stored energy and helicity, maintaining a recurrent eruptive cycle until the available free magnetic energy is significantly reduced.

\section{Conclusions}\label{sec:conc}

This work examined the onset and evolution of jets that follow from the interaction between an emerging toroidal flux tube and an oblique magnetic field in the solar atmosphere. Our study confirmed previous findings about toroidal flux tube emergence \citep{Hood2009,Mac2009}, and expanded these studies by including the ambient field and by running for a longer time.

The initial phases of the simulation are characterized by magnetic flux emerging and expanding into the atmosphere. Soon after, magnetic reconnection between the emerging magnetic flux and the ambient coronal field generates the first ``standard" jet. After the onset of the reconnection jet, a series of six recurrent eruption-driven jets was observed (see Figure \ref{fig5_new}), when the kinetic energy and the temperature showed local peaks during their evolution.

In order to define the start and end of the magnetic flux emergence phase we tracked the temporal evolution above and below the photosphere of various magnetic flux components (see Figure \ref{fig_fl}a). We found that the vertical and axial flux stop emerging at around $t=110$~min, however the azimuthal flux continues until $t=175$~min. In addition, we noticed that the maximum vertical velocity at the photosphere (see Figure \ref{fig_fl}b) decreases rapidly after $t=110$~min, further indicating the decline of flux emergence.

Following the sunspots emergence, magnetically-induced shearing along the PIL lead to the buildup of free magnetic energy and relative helicity inside the coronal volume (see Figure \ref{fig_esh}).
The examination of the evolution of free energy and relative helicity revealed that both quantities respond to the production of the jets by decreasing in absolute values. They rebuild however afterwards and thus maintain the recurrent activity. Interestingly, the eruptive activity continues even after the end of magnetic flux emergence to the solar atmosphere as there is still enough free energy and helicity in the coronal volume.
A similar behavior of recurrent blowout jets was recently found by \cite{Zhou2025} in blowout jet observations from the NOAA active region 13078 in 2022.
These authors found that the evolution of helicity was steadily increasing before declining after the eruption and then increasing again before the next eruption.
This is in agreement with our numerical model, although further study is necessary.
The eruptive activity in our model stops abruptly when the free energy and the relative helicity in the corona decrease significantly and saturate to constant values, which leads to the end of the eruption cycle. Similar conclusions hold also for the average magnetic tension along the axis of the toroidal tube.

Our work can be expanded in a number of ways. A possible future study could be to perform a parametric study in order to examine how the various parameters of the initial toroidal flux tube, such as its twist or magnetic field strength, affect the frequency and nature of the produced eruptions. Another possibility would be to investigate the exact mechanisms that are responsible for generating the eruptive jets during and after the end of flux emergence. The study of the toroidal geometry for the flux tubes in flux emergence experiments presents a promising alternative to the standard cylindrical case.
%


\section*{acknowledgments}
\nolinenumbers
We would like to thank the referee for very helpful and constructive comments that helped to improve our manuscript. This research has been supported by the European Research Council through the Synergy Grant No. 810218 (“The Whole Sun”, ERC-2018-SyG). The simulation demonstrated in this study was supported by computing time awarded from the National Infrastructures for Research and Technology S.A.(GRNET S.A.) in the National HPC facility---ARIS---under the project name MFES.

\bibliography{sample631}{}
\bibliographystyle{aasjournal}



\end{document}